\documentclass[aps,a4paper, preprint, superscriptaddress,preprintnumbers,floatfix,
nofootinbib,amsmath,amssymb]{revtex4}
\usepackage{graphicx}
\usepackage{epsfig}
\usepackage{amsmath}
\usepackage{amsfonts}
\usepackage{amssymb}
\usepackage{url}
\usepackage{subfigure}
\usepackage{hhline}
\usepackage{bm}
\usepackage{cancel}
\usepackage[toc,page]{appendix}
\usepackage{comment}
\usepackage{hyperref}
\usepackage{amsmath}
\usepackage[usenames]{color}
\usepackage{minitoc}
\usepackage{tabulary}
\usepackage{array}
\usepackage{setspace}
\usepackage{enumerate}
\usepackage{acronym}
\usepackage{multirow}
\usepackage{url}
\usepackage{xcolor,listings} 
\usepackage{natbib}
\begin{document}
\title{Neutrino Mass and its Impact on Gravitational Waves from Domain Wall Collision }

\author{Victoria Puyam}
\email{victoria@tezu.ernet.in}
 \affiliation{Department of Physics, Tezpur University, Tezpur-784028, India}  
\author{Mrinal Kumar Das}
\email{mkdas@tezu.ernet.in}
 \affiliation{Department of Physics, Tezpur University, Tezpur-784028,India}

\begin{abstract}
 The $A_{4} \times Z_{4}$ symmetry models are constructed to study neutrino masses and mixings, as well as the gravitational-wave spectrum from domain-wall annihilation. The first neutrino mass model is constructed with the flavon's vacuum expectation value and alignment obtained from the self-interacting potential terms, while the second model uses a new vacuum expectation value arising from a potential containing both self-interaction and mixed terms. The resulting neutrino mixing patterns for both models are in good agreement with current neutrino oscillation data with different mixing values. Further, the flavon mixing terms lift the vacua degeneracy that often shows in the spontaneous symmetry breaking of the discrete symmetry. These mixing terms and the modified neutrino mass matrix of the second model are considered to produce the necessary bias for analysing the gravitational waves spectrum. The spectrum predicted by the model could be detected by current and near-future experiments when the flavons have the vacuum expectation value of $10^4$ TeV.


\end{abstract}

\maketitle
\section{Introduction}
The Standard Model (SM) of particle physics is the most successful theory that explains the behavior and interaction of most of the fundamental particles in nature. However, it has some shortcomings as it fails to explain the dark matter, dark energy, baryon asymmetry of the universe, the finite masses of the neutrinos and their flavor mixing as observed from neutrino oscillation experiments \cite{KamLAND:2002uet, SNO:2002tuh, Super-Kamiokande:1998kpq, DoubleChooz:2011ymz, Lasserre:2012ax, Ling:2013fta, McDonald:2016ixn}, the mass hierarchy among different generations of fermions, and the nature of neutrinos, whether Dirac or Majorana, and larger leptonic mixing compared to quark mixing, to name a few. 

One tackles the issues of flavor dynamics such as lepton flavor and their mass spectra (the mechanism behind their origin), non-trivial flavor mixing in the leptonic sector, CP violation, etc. by using flavor symmetries. Flavor symmetries are often investigated using group theories. These groups can be either Abelian or non-Abelian, and they can also be continuous or discrete \cite{Altmannshofer:2024hmr, King:2013eh}. Among these groups, the non-Abelian discrete symmetry group \cite{Aslam:2026zrw, Chulia:2025nsa, CarcamoHernandez:2022bka, Thapa_2023} and, recently, their modular variants \cite{Pathak:2026ezl, Nomura:2023kwz,deMedeirosVarzielas:2022ihu, Kalita:2026zhm} have gained interest, as they can accommodate neutrino masses and flavor mixing quite naturally.

In the model employing flavor symmetry, the symmetry is often broken spontaneously by a Higgs singlet scalar called a flavon. The number of flavons that can be added to the flavor symmetry models is not restricted by the gauge symmetries. These flavons can self-interact and can also cross-interact depending on the charge assigned to them under the symmetry group of interest. The direct search for flavons at a low energy scale can be tested at the LHC \cite{Heinrich_2019}. Neutrino experiments and rare decay searches can also be used to test flavor symmetry models. However, in most models, the flavor symmetry is assumed to be broken at an ultra-high scale, and signatures arising from it, except for lepton mixing, are often overlooked.

In discrete symmetry cases, whenever flavons spontaneously break the discrete symmetries, degenerate vacua separated by energy barriers form \cite{Gouttenoire_2025, King_2024, jueid2023cosmologicaldomainwallsbreaking}. These energy barriers lead to the formation of cosmological objects called domain walls \cite{Zeldovich:1974uw}. These domain walls are a major shortcoming of discrete symmetry models, as they are not observed and are in disagreement with cosmology if they are stable \cite{Gelmini:2020bqg}.The

annihilation of these domain walls would produce gravitational waves (GWs)  that might be detectable in the near future.  One interesting solution for possibly detectable GWs from the annihilation of these domain walls is the addition of a bias term to lift the vacuum degeneracy. These biases can be introduced as higher-dimensional operators originating from quantum gravity \cite{jueid2023cosmologicaldomainwallsbreaking}, anomalous symmetries to partially lift the degeneracy \cite{CHIGUSA2019249}, etc.  In the A4 symmetry models, an explicit breaking term can be introduced by modifying the right-handed Majorana mass term \cite{Gelmini:2020bqg}, and it can also act as a source of non-zero $\theta_{13}$ and CP violation. In this model, we study bias coming from trilinear terms and flavon cross couplings, as in Ref. \cite{chen2026gravitationalwavesa4neutrino}. Here, the bias is from two sources: (1) the coupling of the flavons, including the $A_{4}$ singlet added to the model, which partially lifts the degeneracy, and  (2)  the small bias from modifying the Majorana mass term resulting from the new vacuum expectation values (vevs) arising from the first case. In this work, we also construct two simple neutrino mass models with and without flavon mixing terms to study the difference in mixing patterns. The GW spectrum is studied in the model where mixing terms are present, along with the new structure of the Majorana neutrino mass matrix resulting from it.

The paper is organized as follows. In Section II, we give a short review of domain walls and the derivation of surface tension with a $Z_{2}$ toy model. Section III deals with the flavon potentials, the vacuum degeneracy, and the bias term necessary for lifting the degeneracy.  Two neutrino mass models with and without flavon cross-coupling terms contribution are studied in Section IV. In Section V, we provide the results of our numerical analysis for neutrino masses and mixing and production of GWs using the parameters from the flavon mixing. The summary and conclusion are given in Section VI.

 \section{Evolution of Domain Walls}
 Cosmic domain walls are topological defects separating different degenerate vacua of the potential. They are formed by the spontaneous breaking of discrete symmetries. The formation of domain walls in the early Universe is not accepted, as the energy density of the walls could dominate the total energy density of the universe, producing power law inflation \cite{Zeldovich:1974uw}. So, the walls need to disappear early enough to prevent the wall energy domination. The walls can annihilate early due to the small energy difference between the minima (bias), as suggested by \cite{Zeldovich:1974uw, PRESKILL1991207}.

To study the breaking and bias, one can consider a toy model in $Z_{2}$ symmetry. The Lagrangian of the real scalar field $\phi$ under $Z_{2}$ symmetry is considered as below 
 \begin{equation}
 L=-\frac{1}{2}\partial^{\mu}\phi\partial_{\mu}\phi-\frac{\lambda}{4}(\phi^2-v^2)^2.
 \label{t1}
 \end{equation}
 The potential has two minima at $<\phi>=\pm v$, and the barrier height is $\lambda v^4/4$. The wall width tends to make the wall thinner while the gradient term tends to make it wider. The gradient term for the model (see Ref \cite{PhysRevD.39.1558}) is given below  

\begin{equation}
\Delta=(\sqrt{\lambda /2}v)^{-1}
\end{equation}

The surface tension, which is equal to the energy per unit area of the wall in its rest frame, is obtained by integrating the 00 component of the wall stress-energy as 

\begin{equation}
\sigma=\frac{2\sqrt{2}}{3}\lambda^{1/2}v^3\equiv f_{\sigma}v^3, 
\end{equation}

 where $f_{\sigma}$ is a dimensionless, real, positive constant and model-dependent. Assuming that the bias term lifts the degeneracy of the vacua, then

\begin{equation}
V_{bias}\equiv \epsilon_{b}v^4,
\end{equation}

where $ \epsilon_{b}$ is the  dimensionless real positive constant and $\epsilon_{b}<<1$. The wall tension $\sigma$ and potential bias term $V_{bias}$ characterise the wall dynamics. 

The potential in Eqn (\ref{t1}) is valid only at temperatures below the phase transition critical temperature $T_{c}$,  which is found to be $T_{c}=2v$ for the toy model \cite{PhysRevD.39.1558}. Here, we assume a radiation dominated universe before the phase transition as well as after it. Thermal fluctuations in the field become large near $T_{c}$ and the regions fluctuate between the two minima. As the temperature goes down below the phase transition temperature $T_{c}$, the fluctuations become progressively rarer due to the increase in the barrier between the two vacua. They become exponentially suppressed, and the patches of the vacua become fixed.

The wall dynamics is governed by surface tension and frictional force. In the scaling regime,  the frictional force is negligible, and the dominant force is the surface tension. Here, the average radius of curvature of the wall is comparable to the Hubble Radius at which $R\sim H^{-1}\sim t$ \cite{Hiramatsu_2014}. The pressure due to tension is obtained as

\begin{equation}
p_{T}\approx \frac{\sigma}{R}=\frac{\sigma}{t}
\end{equation}

The volume pressure due to the energy bias also governs wall dynamics by accelerating the walls toward the false vacuum and converting the false vacuum into true vacuum. It is given by the equation

\begin{equation}
p_{V}\approx V_{b}=\epsilon_{b}v^4
\end{equation}

The domain walls annihilate when the volume pressure becomes comparable to the tension pressure, and from this condition one can calculate the annihilation time as

\begin{equation}
t_{ann}=\frac{\sigma}{V_{b}}
\end{equation}

Further, the upper and lower bounds on the bias are also calculated  in Ref \cite{Gelmini:2020bqg}, and the equation is mentioned below 

\begin{equation}
10^{-25}\left(\frac{\text{TeV}}{v}\right)<\frac{\epsilon{b}}{f_{\sigma}}<10^{-15}\left(\frac{v}{\text{TeV}}\right)
\end{equation}  
\section{Scalars and their Interactions \label{s1}}

In this model, we consider three SM singlets, namely $\phi_{1}$, $\phi_{2}$, and $\chi$. They are assinged with $A_{4}$ charges as $3$, $3$ and $1$, respectively. These flavons, after spontaneous symmetry breaking of the $A_{4}$ group, will acquire non-trivial vacua and preserve some residual symmetries. The scalar potential of the flavons consists of the self-interaction terms and the mixed terms under the $A_{4}$ symmetry. The mixed terms introduce a small perturbation to the vevs of the original $\phi_{1}$ and $\phi_{2}$, which in turn change neutrino mass. The resultant neutrino mass can give a small bias necessary to produce GW from domain wall collision. 

Here, prior to introducing the mixed terms, we study the vevs of the  $\phi_{1}$ and $\phi_{2}$ from their tree-level potentials, respectively. The tree level potential of $\phi_{1}$ is 
\begin{equation}
V_{Tree}(\phi_{1})=\frac{1}{2}\mu_{\phi_{1}}^{2}(\phi_{1}\phi_{1})+\frac{1}{4}(g_{1}(\phi_{1}\phi_{1})^2+g_{2}(\phi_{1}\phi_{1})_{1^{'}}(\phi_{1}\phi_{1})_{1^{''}}+g_{3}(\phi_{1}\phi_{1})_{3}(\phi_{1}\phi_{1})_{3})+g_{4}(\phi_{1}\phi_{1}\phi_{1})
\label{e1}
\end{equation}

where, $\mu_{\phi_{1}}^{2}$ is mass like term, and $g_{1}$, $g_{2}$, $g_{3}$ and $g_{4}$ are free coupling terms. In MR basis, Eq.~(\ref{e1}) can be written as 
\begin{align}
V_{Tree}(\phi_{1})=&\frac{1}{2}\mu_{\phi_{1}}^{2}I_{1}(\phi_{1})+\frac{f_{1}}{4}I_{1}^{2}(\phi_{1})+\frac{f_{2}}{4}I_{2}^{2}(\phi_{1})+f_{3}\phi_{1_{1}}\phi_{1_{2}}\phi_{1_{3}}
\end{align}

with $f_{1}=g_{1}+g_{2}$, $f_{2}=\frac{3(f_{3}-f_{2})}{2}$, $g_{3}=\sqrt{3}f_{4}$, and 
\begin{align}
I_{1}(\phi_{1})=&(\phi_{1_{1}}^{2}+\phi_{1_{2}}^{2}+\phi_{1_{3}}^{2})\nonumber \\I_{2}(\phi_{1})=&(\phi_{1_{1}}^{2}\phi_{1_{2}}^{2}+\phi_{1_{1}}^{2}\phi_{1_{3}}^{2}+ \phi_{1_{2}}^{2}\phi_{1_{3}}^{2}).
\label{e5} 
\end{align}
The parameters $\mu_{\phi_{1}}^{2}<0$, $f_{1}>0$, $f_{2}>0$, and $f_{3}>0$ are considered to ensure the spontaneous breaking of the $A_{4}$ symmetry to non-trivial vacuum. After minimization of  $V_{Tree}(\phi_{1})$, the degenerate vacua obtain are given below:

\begin{align}
\left\{
\begin{pmatrix}
1\\1\\1
\end{pmatrix},\begin{pmatrix}
1\\-1\\-1
\end{pmatrix},
\begin{pmatrix}
-1\\1\\-1
\end{pmatrix},
\begin{pmatrix}
-1\\-1\\1
\end{pmatrix}
\right\} v(\phi_{1}^{-})\nonumber \\ 
\nonumber \\ 
\left\{
\begin{pmatrix}
-1\\1\\1
\end{pmatrix},\begin{pmatrix}
1\\-1\\1
\end{pmatrix},
\begin{pmatrix}
1\\1\\-1
\end{pmatrix},
\begin{pmatrix}
-1\\-1\\-1
\end{pmatrix}
\right\} v(\phi_{1}^{+}).
\label{e3}
\end{align}

 The set of degenerate vacua in Eqn.(\ref{e3}) preserve $Z_{3}$ residual symmetry with $v(\phi_{1}^{\pm})=\frac{\mu_{\phi_{1}}}{\sqrt{3f_{1}+2f_{2}}}(\sqrt{1+(\frac{f_{3}}{\mu_{\phi_{1}}\sqrt{3f_{1}+2f_{2}}}})^2\pm \mu_{\phi_{1}}\sqrt{3f_{1}+2f_{2}})$.The stability of the different sets of degenerate vacua preserving other residual symmetry arising from  different signs of the parameters can be tested by evaluating the eigenvalues of the Hessian mass matrix given by

\begin{equation}
(M_{\phi_{1}}^2)_{ij}=\frac{\partial^{2}V_{Tree}(\phi_{1})}{\partial \phi_{1_{i}}\partial \phi_{1_{j}}}|_{<\phi_{1}>}.
\label{e4}
\end{equation} 

After calculating Eqn. (\ref{e4}) using the parameters that yield the set of degenerate vacua in Eqn. (\ref{e3}), and subsequently diagonalising the result, one can derive the mass eigenvalues \( m_{\phi_{1_{1}}}^2 \), \( m_{\phi_{1_{2}}}^2 \), and \( m_{\phi_{1_{3}}}^2 \), where 

\begin{align}
m_{\phi_{1_{1}}}^2=&2\mu_{\phi_{1}}\left(1+(\frac{f_{3}}{\mu_{\phi_{1}}\sqrt{3f_{1}+2f_{2}}})^2\pm \frac{f_{3}}{\mu_{\phi_{1}}\sqrt{3f_{1}+2f_{2}}}\sqrt{(1+\frac{f_{3}}{\mu_{\phi_{1}}\sqrt{3f_{1}+2f_{2}}})^2}\right)
\nonumber \\
m_{\phi_{1_{2}}}^2=&m_{\phi_{1_{3}}}^2=\frac{g_{3}-g_{2}}{2g_{2}+g_{1}+g_{3}}\mu_{\phi_{1}}[1-2\frac{f_{3}}{\mu_{\phi_{1}}\sqrt{3f_{1}+2f_{2}}}(\frac{f_{3}}{\mu_{\phi_{1}}\sqrt{3f_{1}+2f_{2}}}\pm \sqrt{(1+\frac{f_{3}}{\mu_{\phi_{1}}\sqrt{3f_{1}+2f_{2}}})^2})\nonumber \\ &(1+2\frac{f_{1}+f_{2}}{f_{2}-f_{3}})].
\end{align}

Further, we discuss the tree level potential of $\phi_{2}$, and it takes the form
\begin{equation}
V_{Tree}(\phi_{2})=\frac{1}{2}\mu_{\phi_{2}}^{2}(\phi_{2}\phi_{2})+\frac{1}{4}(h_{1}(\phi_{2}\phi_{2})^2+h_{2}(\phi_{2}\phi_{2})_{1^{'}}(\phi_{2}\phi_{2})_{1^{''}}+h_{3}(\phi_{2}\phi_{2})_{3}(\phi_{2}\phi_{2})_{3})
\label{e9}
\end{equation}
where, $\mu_{\phi_{2}}^{2}$, $h_{1}$, $h_{2}$ and $h_{3}$ are coupling constants and it can be also written in terms of $I_{1}$ and $I_{2}$  defined in Eqn. (\ref{e4}) as
\begin{equation}
V_{Tree}(\phi_{2})=\frac{1}{2}\mu_{\phi_{2}}^{2}I_{1}(\phi_{2})+\frac{k_{1}}{4}I_{1}^{2}(\phi_{2})+\frac{k_{2}}{4}I_{2}^{2}(\phi_{2}).
\end{equation}

where, $\mu_{\phi_{2}}^{2}$, $k_{1}$, and $k_{2}$ are coupling constants with $k_{1}=h_{1}+h_{2}$ and $k_{2}=\frac{3(h_{3}-h_{2})}{2}$ . Here,  $\mu_{\phi_{2}}^{2}<0$, $k_{1}>0$, and $k_{2}>0$ are considered to ensure the spontaneous symmetry breaking of the $A_{4}$ to non-trivial vacuum. The potential  $V_{Tree}(\phi_{2})$ are further minimized and it gives the following set of the vevs
\begin{equation}
\left\lbrace
\begin{pmatrix}
1\\0\\0
\end{pmatrix},
\begin{pmatrix}
0\\1\\0
\end{pmatrix},
\begin{pmatrix}
0\\0\\1
\end{pmatrix},
\begin{pmatrix}
-1\\0\\0
\end{pmatrix},
\begin{pmatrix}
0\\-1\\0
\end{pmatrix},
\begin{pmatrix}
0\\0\\-1
\end{pmatrix}
  \right\rbrace v_{\phi_{2}}
  \label{e6}
\end{equation}

where $v_{\phi_{2}}=\sqrt{-\mu_{\phi_{2}}^2/k_{1}}$ and assumed to be positive. The set of vacua obtained in Eqn.(\ref{e6})preserves $Z_{2}$ residual symmetry. After evaluating the mass eigenvalues of the Hessian mass matrix given in Eqn. (\ref{e4}) with $V_{Tree}(\phi_{2})$ for scalar $\phi_{2}$, and diagonalising it, we obtain

\begin{align}
m_{\phi_{2_{1}}}^2= 2k_{1}v_{\phi_{2}}^2, m_{\phi_{2_{2}}}^2=m_{\phi_{2_{3}}}^2=k_{2}v_{\phi_{2}}^2
\end{align}

Here, both the scalars $\phi_{1}$ and $\phi_{2}$ are taken as psuedo-scalar as in \cite{Pascoli_2016}. The above calculations are in the MR basis; however, in our work, the neutrino mass models are to be constructed in the Altarelli and Feruglio (AF) flavor basis. The related calculations on the AF basis are given in Appendix \ref{b2}. One can find the detailed calculation on both the MR basis and the AF basis in \cite{Gelmini:2020bqg,Pascoli_2016,Pascoli_2016a, chen2026gravitationalwavesa4neutrino}. 

\subsection{Flavon Mixing Terms and its Impact on Vacuum Expectation Values and Alignment}
 As mentioned earlier, the full potential of the scalar sector interaction consists of both $V_{Tree}$ and flavon mixing terms. The contribution from the mixing terms is considered very small to preserve the $Z_{3}$ and $Z_{2}$ residual symmetries in the leading order. Since our neutrino mass model is to be constructed in the AF basis, we express the potential in the AF basis, and the $A_{4}$ symmetry is augmented with auxiliary $Z_{4}$ symmetry. Now, the scalars $\phi_{1}$,$\phi_{2}$  and $\chi$ have  $A_{4} \times Z_{4}$ charges $(3,0)$, $(3,2)$, and $(1,2)$, respectively and the potential due to mixing terms ($V(\phi_{1},\phi_{2},\chi)$) becomes
\begin{align}
V(\phi_{1},\phi_{2},\chi)=& \frac{1}{2}\epsilon_{1}(\phi_{1}\phi_{1})_{1} (\phi_{2}\phi_{2})_{1} +\frac{1}{4}\epsilon_{2}(\phi_{1}\phi_{1})_{1^{'}} (\phi_{2}\phi_{2})_{1^{''}}+\frac{1}{4}\epsilon_{2}^{*}(\phi_{1}\phi_{1})_{1^{''}} (\phi_{2}\phi_{2})_{1^{'}}+\frac{1}{2}\epsilon_{3}(\phi_{1}\phi_{1})_{3} (\phi_{2}\phi_{2})_{3}\nonumber \\ &+\frac{1}{3}\epsilon_{4}[(\phi_{2}\phi_{2})_{3} \phi_{1}]_{1},\frac{\epsilon_{5}}{4}(\phi_{1}\phi_{1})_{1}\chi^2 + \frac{\epsilon_{6}}{4}(\phi_{2}\phi_{2})_{1}\chi^2 +\frac{\epsilon_{7}}{4}(\phi_{1}\phi_{2})_{1}\chi+\frac{\epsilon_{8}}{4}(\phi_{1}\phi_{1}\phi_{2})_{1}\chi+\nonumber \\ & \frac{\epsilon_{9}}{4}(\phi_{1}\phi_{1}\phi_{1})_{1}\chi
\end{align} 

where $ \epsilon_{1}$, $\epsilon_{2}$, $\epsilon_{2}^{*}$, $\epsilon_{3}$, $\epsilon_{4}$, $\epsilon_{5}$, $\epsilon_{6}$, $\epsilon_{7}$,$\epsilon_{8}$, and $\epsilon_{9}$ are coupling coefficients and they are assumed  to be  small to preserve the residual symmetries in the leading order. Here, $V_{tree}(\chi)=1/2 \mu_{\chi}^2 \chi^2+(r_1/4 )\chi^4$  with $\mu_{\chi}^2$ and $r_1$ are coupling constants with vev $v=(-mt/r1)^{1/2}$ In the AF basis, the vev of $\phi_{1}$ and $\phi_{2}$ are considered as 
\begin{equation}
\phi_{1}=\begin{pmatrix}
1\\0\\0
\end{pmatrix},v_{\phi_{1}}\ \text{and} \ \phi_{2}=\begin{pmatrix}
1\\1\\1
\end{pmatrix}v_{\phi_{2}}.
\label{e7}
\end{equation}

The vacuum in Eqn. (\ref{e7}) is assumed to be the true vacuum, and the small deviation from this vacuum due to the mixed terms is found to be
\begin{equation}
\phi_{1}=\begin{pmatrix}
v_{\phi_{1}}+ \epsilon_{0}\\  \epsilon_{l}\\  \epsilon_{l}^{*}
\end{pmatrix},\ \text{and} \ \phi_{2}=\begin{pmatrix}
1 + \epsilon_{N}\\ 1+ \epsilon_{n}\\ 1+ \epsilon_{n}
\end{pmatrix}v_{\phi_{2}}\  \text{and}\ =u-u_{o}.
\label{e7}
\end{equation}
Here, $\epsilon_{0}$ and $u_{o}$ can be reabsorbed into $ v_{\phi_{1}}$ and $u$, respectively, and,
\begin{align}
 \epsilon_{l}=&-\frac{(3 \epsilon_{2}v_{\phi_{2}}-\epsilon_{8}v)v_{\phi_{2}}}{2(v_{\phi_{1}}g_{1}+v_{\phi_{1}}g_{2})+v_{\phi_{1}}(3v_{\phi_{1}}g_1+2v_{\phi_{1}}g2+6g_3)^2 -2g_{4})}, \nonumber \\
\epsilon_{l}^*=&-\frac{(3 \epsilon_{2}^*v_{\phi_{2}}-\epsilon_{8}v)v_{\phi_{2}}}{2(v_{\phi_{1}}g_{1}+v_{\phi_{1}}g_{2})+v_{\phi_{1}}(3v_{\phi_{1}}g_1+2v_{\phi_{1}}g2+6g_3)^2 -2g_{4})},\nonumber \\
\epsilon_{N}=&\phi_{2}\epsilon_{n}
\ \text{and}\ \epsilon_{n1}=\epsilon_{n}-(\frac{\phi_{1}(18\phi_{2}\epsilon{4}+\phi_{2}\phi_{1}(3\epsilon_{1}+18\epsilon_{3}+4v\epsilon_{7})+3\phi_{1}v\epsilon_{8}}{3\phi_{2}^3(4h_1-5h_2+36h_3)})\phi_{2},\nonumber \\
\text{where}\ &\epsilon{n}=(\phi_{2} (96 \phi_{1}(h_1+h_2\epsilon_4-3(4h_1-5h_2+36 h_3)v^2\epsilon_6+\phi_{1}^2(21 h_2\epsilon_{1}-36h_3\epsilon_{1}+96h _2\epsilon_{3}+\nonumber \\
&28 h_2 v\epsilon_{7}-48 h_3 v\epsilon_{7} + 4h_1(3 \epsilon_{1} +24\epsilon_{3}+4 v \epsilon_ {7})))+3 \phi_{1} ^2 (4 h_1+7h_2-12 h_3)v\epsilon_{8})/(48 \nonumber \\&\phi_{2}^2 (h_1+h_ 2)
(4h_1-5h_2+36h_3)).\nonumber \\
\end{align} 

  Here, the potentials $V_{Tree}(\phi_{1})$ and $V_{Tree}(\phi_{2})$ both have same interactions as Eqn. (\ref{e1}) and  Eqn. (\ref{e9}) in the AF basis, but the tensor multiplication shall follow Eqn. (\ref{af}). The detailed calculation can be found in \cite{Gelmini:2020bqg,chen2026gravitationalwavesa4neutrino}. The bias term $(V(\phi_{1},\phi_{2}, \phi_{3}$ can partially lift the degeneracy but cannot differentiate $Z_{2}$ degeneracy in $\phi_{2}$. This degeneracy will be eventually lifted by $V_{loop}$ due to the neutrino mass matrix. 
\section{Neutrino Mass Model \label{n1}}
In this work, we considered a model whose field contents are similar to the original AF model. In that model, the neutrino mass model preserves Tribimaximal mixing (TBM), and the charged lepton mass matrix is diagonal when the vevs are $(1,1,1)^{T}$ and $(1,0,0)^{T}$, respectively. The introduction of new vevs alignment due to flavon mixing terms gives deviation from both the TBM and the diagonal charged lepton mass matrix. In this model, we will consider both the new vevs alignment due to flavon mixing terms and the neutrino mass matrix arising from both the non-trivial Dirac mass matrix and Majorana mass matrix in the type-I seesaw mechanism. The field contents of the model are given below 
\ref{t1}
\begin{table}[h!]
\small
\begin{center}
\begin{tabular}{|c|c|c|c|c|c|c|c|c|c|c|c|c|c|c|c|}
\hline 
Fields & $l$ & $e^{c}$ & $\mu^{c}$  & $\tau^{c}$& $\nu^{c}$  & $H_{u,d}$ & $\phi_{1}$& $\phi_{2}$ & $\chi$ \\ 
\hline 
$A_{4}$ & 3& 1& $1^{''}$ &  $1^{'}$&3& 1 & 3 & 3& 1 \\ 
\hline 
$Z_{4}$ &1 & 3 & 3 &3& 1 &0 & 0 & 2 &2\\
\hline
$SU(2)_{L}$& 2& 1&1&1&1& 2&1&1&1\\
\hline
\end{tabular}
\caption{\footnotesize{Transformation properties of various fields under $A_{4}\times Z_{3} \times Z_{2}\times SU(2)_{L}$ group. Here, the $Z_{4}$ charge are in addition notation.}}
\label{t1}
\end{center}
\end{table}

The Yukawa Lagrangian  which are invariant under $A_{4} \times Z_{4}$ group, are given in the equation:
\begin{align}
L_{l}= &\frac{Y_{e}}{\Lambda}(l\phi_{1})_{1}H_{d}e^{c}+\frac{Y_{\mu}}{\Lambda}(l\phi_{1})_{1^{'}}H_{d}\mu^{c}+\frac{Y_{\tau}}{\Lambda}(l\phi_{1})_{1^{''}}H_{d}\tau^{c}+ \frac{ y_{1}}{\Lambda}\chi(lH_{u}\nu^{c})_{1}+ \frac{y_{a}}{\Lambda}\phi_{2}(lH_{u}\nu^{c})_{A}\nonumber \\ &+\frac{y_{b}}{\Lambda}\phi_{2}(lH_{u}\nu^{c})_{S}  +y_{n1}(\nu^{c}\nu^{c})\chi+y_{n2}(\nu^{c}\nu^{c})\phi_{2}+h.c
\label{e8}
\end{align}

After spontaneous symmetry breaking (SSB), the vev of the $< \phi_2 > $ and $ < \phi_2 >$, $<\chi>$, and $h_{u,d}$ are assumed to be  $(1, 1, 1)^{T}v_{\phi_{2}}$ , $ < \phi_2 >= (1, 0, 0)^{T}v_{\phi_{1}}$, $v$, and $u_{u,d}$ respectively. The lepton mass matrices take the forms

\begin{align}
M_{l1}=&\begin{pmatrix}
m_e&0&0\\
0&m_{\mu}&0\\
0&0&m_{\tau}
\end{pmatrix},\ 
M_{D1}=\begin{pmatrix}
a + 2 b& -b + c& -b - c \\
 -b - c& 2 b& a - b + c\\
 -b + c& a - b - c& 2 b
\end{pmatrix}, \nonumber \\ \vspace{0.5cm}
\text{and}\ M_{N1}=&\begin{pmatrix}
A + 2 B & -B & -B\\
 -B& 2 B& A - B\\
 -B & A - B & 2 B
\end{pmatrix}
\end{align}
where  $m_{e}=Y_{e} v_{\phi_{1}} u_{d}/\lambda$,  $m_{\mu}=Y_{\mu} v_{\phi_{1}} u_{d}/\lambda$, $m_{\tau}=Y_{\tau} v_{\phi_{1}} u_{d}/\lambda$, $a=y_{1} v u_{u}/\lambda$, $c=y_{a} v_{\phi_{2}} u_{u}/\lambda$, $b=y_{b} v_{\phi_{2}} u_{u}/\lambda$, $A=y_{n1}v$ and $B=y_{n2}v_{\phi_{2}}$,while $M_{l}$, $M_{D}$, and $M_{N}$ are the charged lepton mass matrix, Dirac mass matrix and the Majorana mass matrix, respectively. The light neutrino mass matrix $(m_{\nu 1})$ is obtained from type I seesaw mechanism as
\begin{equation}
m_{\nu 1}=M_{D1}M_{R1}^{-1}M_{D1}^{T}
\end{equation}
\begin{equation}\label{e10}
=\footnotesize{\frac{1}{M}\begin{bmatrix}
m_{11}& m_{12}& m_{13}\\
m_{21}& m_{22}& m_{23}\\
m_{31}& m_{32}& m_{33}
\end{bmatrix}},
\end{equation}
where, 
\begin{align*}
M=& A^3 - 9 A B^2 \nonumber \\
m_{11}=&(4 a A b + 6 A b^2 + a^2 (A + B))(A - 3 B)-2c^2(A (A + 3 B))\nonumber \\
m_{12}=&m_{21}=(A - 3 B) (-A b (2 a + 3 b) + a^2 B) + 6 A (A b - a B) c + A (A + 3 B) c^2 \nonumber\\
 m_{13}=&m_{31}=(A - 3 B) (-A b (2 a + 3 b) + a^2 B) + 6 A (-A b + a B) c + A (A + 3 B) c^2 \nonumber \\
 m_{23}=&m_{32}= (-2 a A b (A + 6 B) + a^2 (A^2 + A B - 3 B^2) + 
 A (6 A b^2 + 9 b^2 B - 2 A c^2 + 3 B c^2))\nonumber \\
 m_{22}=& a^2 B (2 A + 3 B) + A (-3 b^2 (A + 6 B) - 6 A b c + (A - 6 B) c^2) + 2 a A (2 A b + 3 B (b + c))\nonumber \\
 m_{33}=&(-a^2 B (2 A + 3 B) + 2 a A (2 A b + 3 B (b - c)) + A (-3 b^2 (A + 6 B) + 6 A b c + (A - 6 B) c^2))
\end{align*}

The active neutrino mass matrix obtained in Eqn. (\ref{e10}) can give a deviation from TBM, unlike the original AF model. This model can also explain the current neutrino oscillation data. However, the value of $\theta_{12}$ is centered around $35^{\circ}$, which differs from the neutrino mass matrix constructed using new vevs.  Further, we also study the change in $m_{\nu}$ due to flavon mixing potential terms.  The lepton mass matrices obtained with the corrected vevs found in Eqn. (\ref{e7}) are 
\begin{align*}
M_{l}=&\begin{pmatrix}
Y_{e}u_{d}v_{\phi_{1}}&Y_{\mu}u_{d}\partial \epsilon_{l}&Y_{\tau}u_{d}\partial \epsilon_{l}^{*}\\
Y_{e}u_{d}\partial \epsilon_{l}^*&Y_{\mu}u_{d} v_{\phi_{1}}&Y_{\tau}u_{d}\partial \epsilon_{l}^{*}\\
Y_{e}u_{d}\partial \epsilon_{l}&Y_{\mu}u_{d}\partial \epsilon_{l}&Y_{\tau}u_{d} v_{\phi_{1}}
\end{pmatrix}, 
\end{align*}
\begin{align*}
\tiny M_{D}=\begin{pmatrix}
\frac{y_{1} v u_{u}}{\lambda} + \frac{2y_{b}(1+\epsilon_{n1}) v_{\phi_{2}} u_{u} }{\lambda}& -\frac{y_{b}(1+\epsilon_{n}) v_{\phi_{2}} u_{u} }{\lambda} + \frac{y_{a}(1+\epsilon_{n}) v_{\phi_{2}} u_{u} }{\lambda}&-\frac{y_{b}(1+\epsilon_{n}) v_{\phi_{2}} u_{u} }{\lambda} - \frac{y_{a}(1+\epsilon_{n}) v_{\phi_{2}} u_{u} }{\lambda} \\
 -\frac{y_{b}(1+\epsilon_{n}) v_{\phi_{2}} u_{u} }{\lambda} - \frac{y_{a}(1+\epsilon_{n}) v_{\phi_{2}} u_{u} }{\lambda}& 2\frac{y_{b}(1+\epsilon_{n}) v_{\phi_{2}} u_{u} }{\lambda} & \frac{y_{1} v u_{u}}{\lambda} - \frac{y_{b}(1+\epsilon_{n1}) v_{\phi_{2}} u_{u} }{\lambda} + \frac{y_{a}(1+\epsilon_{n1}) v_{\phi_{2}} u_{u} }{\lambda}\\
 -\frac{y_{b}(1+\epsilon_{n}) v_{\phi_{2}} u_{u} }{\lambda} + \frac{y_{a}(1+\epsilon_{n}) v_{\phi_{2}} u_{u} }{\lambda}& \frac{y_{1} v u_{u}}{\lambda} - \frac{y_{b}(1+\epsilon_{n1}) v_{\phi_{2}} u_{u} }{\lambda} - \frac{y_{b}(1+\epsilon_{n1}) v_{\phi_{2}} u_{u} }{\lambda}& 2 \frac{y_{b}(1+\epsilon_{n}) v_{\phi_{2}} u_{u} }{\lambda}
\end{pmatrix},
\end{align*}
\begin{align}
\text{and}\ M_{N}=&\begin{pmatrix}
\frac{y_{n1} v u_{u}}{\lambda} + \frac{2y_{n2}(1+\epsilon_{n1}) v_{\phi_{2}} u_{u} }{\lambda} & -\frac{y_{n2}(1+\epsilon_{n}) v_{\phi_{2}} u_{u} }{\lambda} & -\frac{y_{n2}(1+\epsilon_{n}) v_{\phi_{2}} u_{u} }{\lambda}\\
 -\frac{y_{n2}(1+\epsilon_{n}) v_{\phi_{2}} u_{u} }{\lambda}& 2 \frac{y_{n2}(1+\epsilon_{n}) v_{\phi_{2}} u_{u} }{\lambda}& \frac{y_{n1} v u_{u}}{\lambda} - \frac{y_{n2}(1+\epsilon_{n1}) v_{\phi_{2}} u_{u} }{\lambda}\\
 -\frac{y_{n2}(1+\epsilon_{n}) v_{\phi_{2}} u_{u} }{\lambda} & \frac{y_{n1} v u_{u}}{\lambda} - \frac{y_{n2}(1+\epsilon_{n1}) v_{\phi_{2}} u_{u} }{\lambda} & 2 \frac{y_{n2}(1+\epsilon_{n}) v_{\phi_{2}} u_{u} }{\lambda}
\end{pmatrix}
\end{align}

The light neutrino mass matrix is similarly obtained from the type I seesaw mechanism and has the same structure of $m_{\nu}$ as in  Eqn. (\ref{e10}) with different values of mass matrix elements.

\subsection{Diagonalization}   
In the first model, the charged lepton mass matrix is diagonal, and the neutrino mass matrix $m_{\nu}$ obtained in Eqn. (\ref{e10}) can be diagonalised as
\begin{equation}
U^{T}m_{\nu}U=\begin{pmatrix}
m_{1}&0&0\\
0&m_{2}&0\\
0&0&m_3
\end{pmatrix}
\end{equation}
where, $U$ is the  $U_{PMNS}$ matrix, and $m_1$, $m_1$ and $m_3$ are mass eigenvalues. However, for the second model, the charged lepton mass matrix and neutrino mass matrix are non-diagonal. So, the $U_{PMNS}$ matrix has contributions from both the charged lepton mass matrix and neutrino mass matrix, and can be written as
\begin{equation}
U_{PMNS}=U_{l}U_{\nu}
\end{equation}
where, $U_{l}$ and $U_{\nu}$ are the unitary matrix that diagonalise the $M_{l}$ and $m_{nu}$, respectively as

\begin{align}
U_{l}^{\dagger}M_{l}^{\dagger}M_{l}U_{l}=\begin{pmatrix}
me^2&0&0\\
0&m_{\mu}^2&0\\
0&0&m_{\tau}^2,
\end{pmatrix} \ \text{and} \
U^{\dagger}_{\nu}m_{\nu}U_{\nu}=\begin{pmatrix}
m_{1}&0&0\\
0&m_{2}&0\\
0&0&m_3
\end{pmatrix}
\end{align}

Here, $U_{l}$ has the following form
\begin{equation}
\begin{pmatrix}
1&- \epsilon_{1}&- \epsilon_{1}^*\\
- \epsilon_{1}^*&1&- \epsilon_{1}\\
 \epsilon_{1}& \epsilon_{1}^*&1.
\end{pmatrix}
\end{equation}
 The unitary matrices $U_{\nu}$ and $U$ are obtained numerically in both models. In both models, most neutrino oscillation parameters are well explained within current bounds, and their predictions are similar. However, an interesting difference can be seen in the $\theta_{12}$ values, where $\theta_{12}$ is centred around $35^{\circ}$ in the model I while in the model II, $\theta_{12}$ is found to below $35^{\circ}$. In addition, the first model will produce a wall-overclosed universe. Still, GW due to domain wall collision that falls within the sensitivity of current and near-future experiments can be expressed by keeping the constants of flavon mixing terms very small. This small contribution gives new degenerate vevs as mentioned in Section \ref{s1}. The new degenerate vevs, when input to the neutrino mass matrix, especially to the Majorana mass matrix, will give different values of neutrino masses. The small difference in the neutrino masses corresponding to the different degenerate vacua split the degenerate vacua and, produce a small bias necessary for producing GW via domain wall collision. 
 
 \subsection{Bias from the Neutrino Mass Matrix} 
 
As mentioned in the previous section, the neutrino mass terms split the degenerate vacua, and the bias is generated at the loop level. The one-loop correction to the potential, especially to $\phi_{2}$  in  Eqn. (\ref{e6}) can be written as

\begin{equation}
 V_{loop}(\phi_{2},\chi)=\frac{1}{64 \pi^{2}}\text{Tr}\{[M_{N}(\phi_{2},\chi)M_{N}^{\dagger}(\phi_{2},\chi)]^2[\text{log}\frac{M_{N}(\phi_{2},\chi)M_{N}^{\dagger}(\phi_{2},\chi)}{\mu^2}-\frac{3}{2}]\}
\end{equation}   

 where $M_{N}(\phi_{2},\chi)$ is the neutrino mass matrix that depends on $\phi_{2}$ and $\chi$. Here, the flavons $\phi_{2}$ and $\chi$ also interact with $M_{D}$, but this contribution is negligible, as it is suppressed by the cut-off scale $\lambda$. The introduction of  $V_{loop}$ splits the degenerate vacua into different energy vacua. From the neutrino mass matrix $M_{N}$, one can see that the corrected vev contribution $\epsilon_{n1}$ and $\epsilon_{n}$ are the source that splits the degenerate vacua. This value is fixed using neutrino oscillation data.

\section{Results}
\begin{figure}[h]
\centering
\subfigure[]{
\includegraphics[width=.40\textwidth]{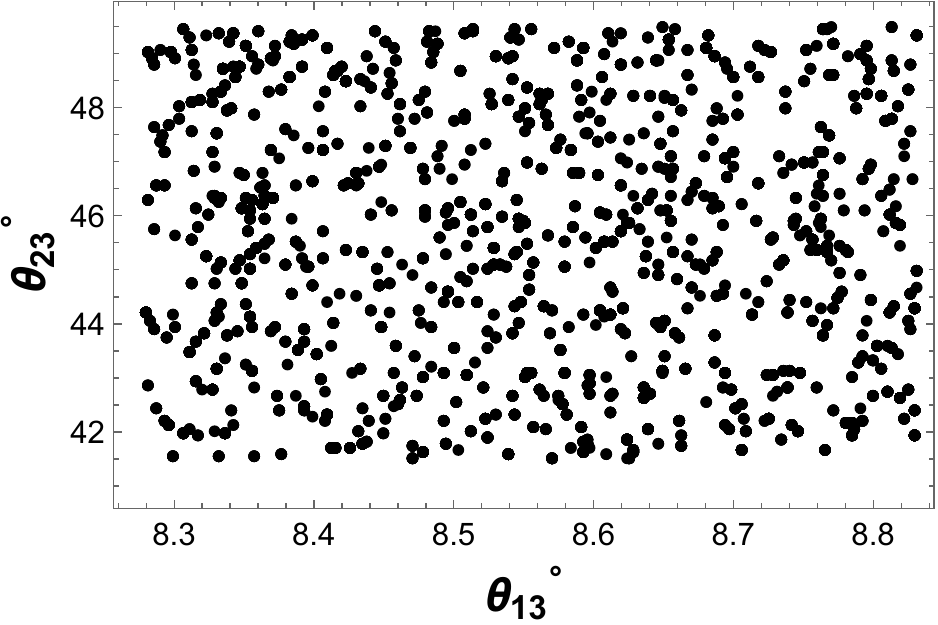}}
\quad
\subfigure[]{
\includegraphics[width=.40\textwidth]{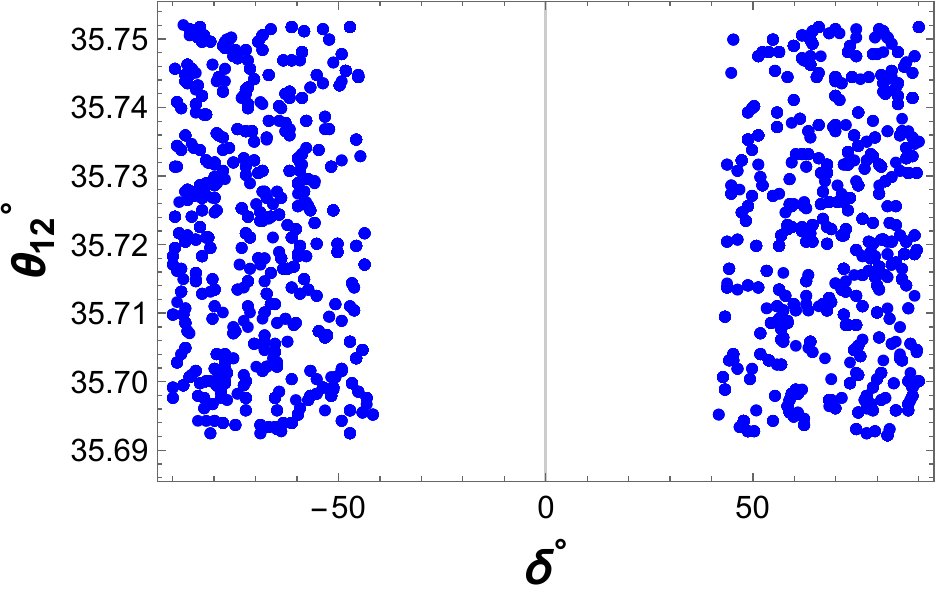}}
\quad
\subfigure[]{
\includegraphics[width=.40\textwidth]{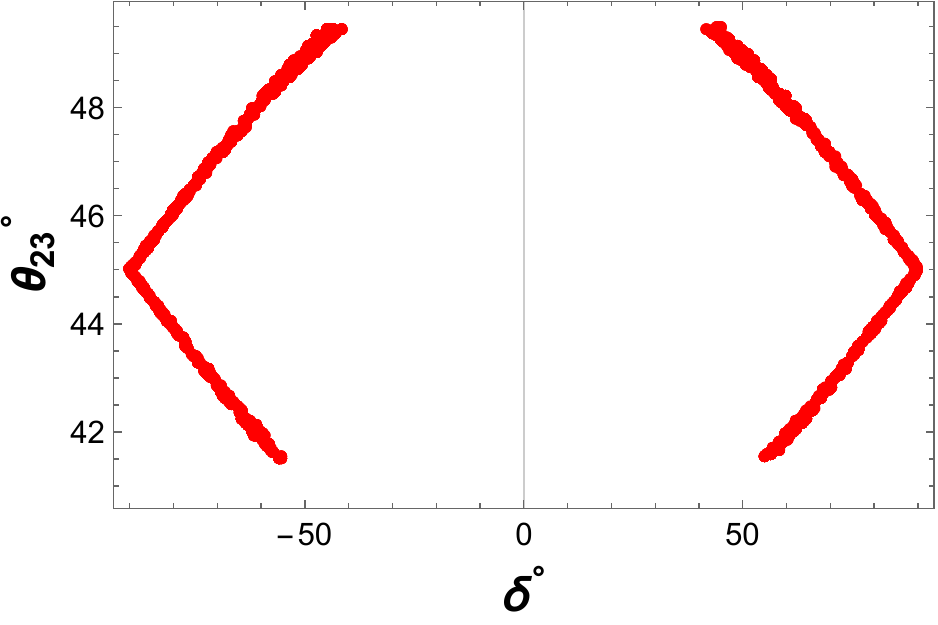}}
\quad
\subfigure[]{
\includegraphics[width=.40\textwidth]{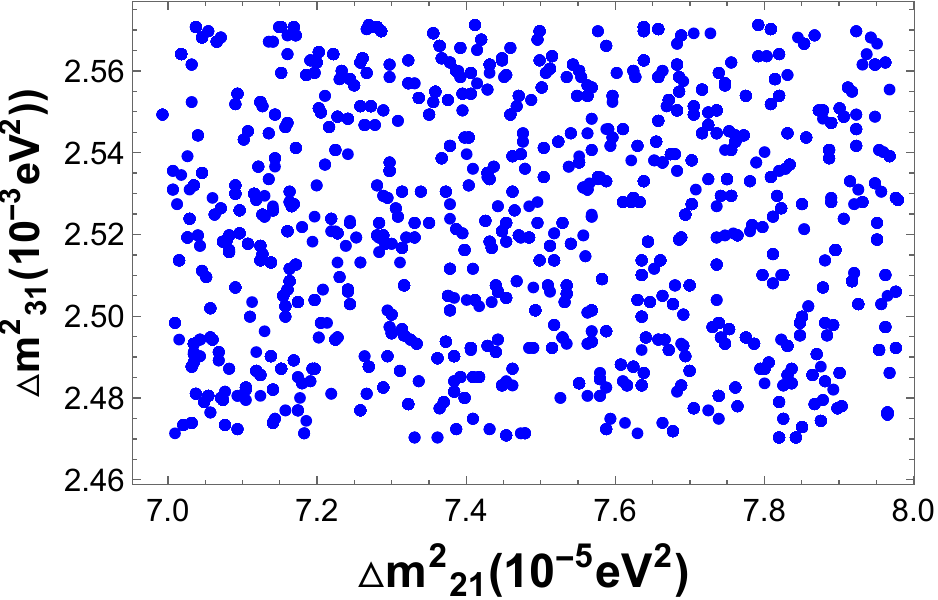}}
\caption{ Plots for the neutrino oscillations parameters for model with no  correction to vevs.}
\label{f0} 
\end{figure}

Following the diagonalization of light neutrino mass matrices obtained in Section \ref{n1}, we can explain the current neutrino oscillation data. In both models, the parameters $A$ and $B$ are considered to be in the ratio $5:1$ to obtain the correct mixing pattern. From the Fig. \ref{f0}, one can see that most of the neutrino oscillation parameters are in good agreement with global fit data \cite{Esteban_2024} with the $\theta_{12}$ center around $35.7^{\circ}$ as the first model with anti-symmetric contribution from Dirac mass terms cannot perturb $\theta_{12}$ values obtain from Tribimaximal mixing. Here, the values of $\theta_{23}$ are also evenly spread throughout the $3 \sigma$ range of the current global neutrino oscillation data.
\begin{figure}[h]
\centering
\subfigure[]{
\includegraphics[width=.40\textwidth]{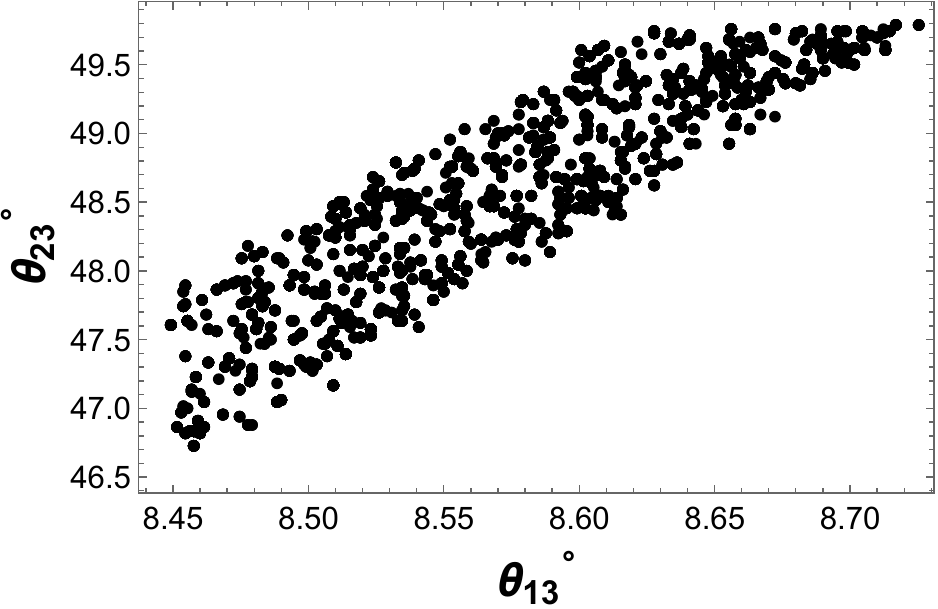}}
\quad
\subfigure[]{
\includegraphics[width=.40\textwidth]{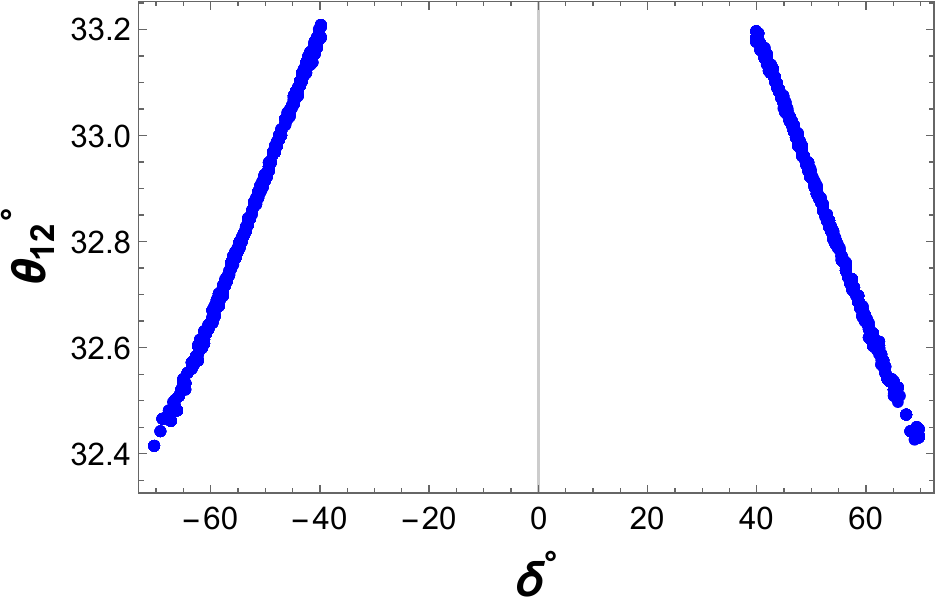}}
\quad
\subfigure[]{
\includegraphics[width=.40\textwidth]{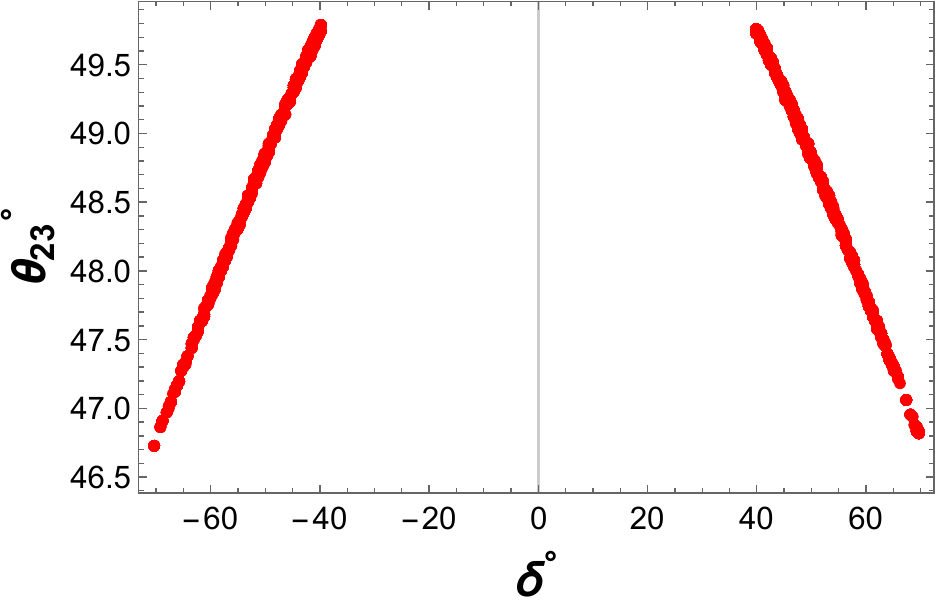}}
\quad
\subfigure[]{
\includegraphics[width=.40\textwidth]{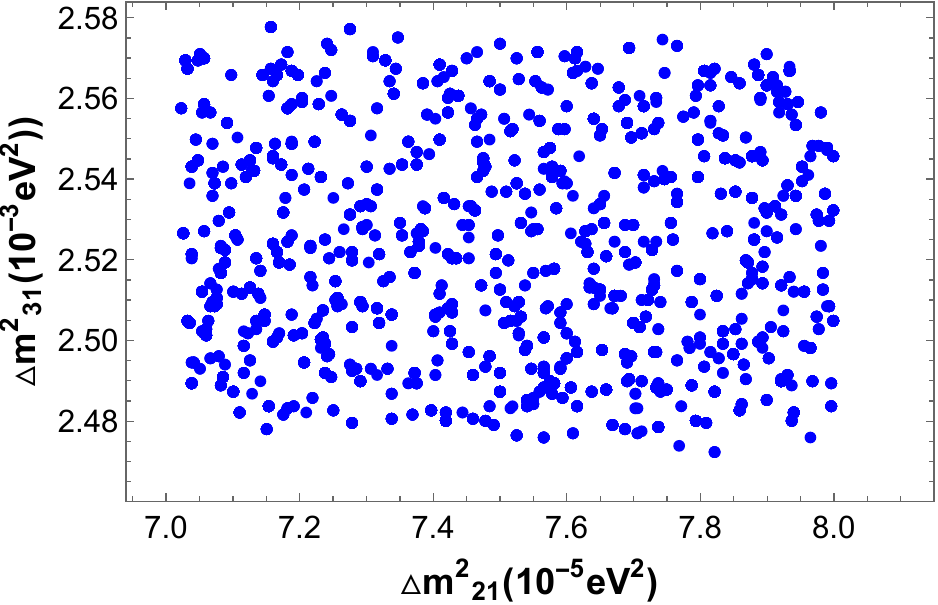}}
\caption{ Plots for the neutrino oscillations parameters for model with vevs correction.}
\label{f1} 
\end{figure}
 
 The plots in Fig. \ref{f1} show the predictions of neutrino oscillation parameters in a model with vev corrections. The oscillation parameters are in good agreement with current data, and most results are similar to the first model. However, introduction of new terms due to flavon mixing in the potential give some interesting changes in the value of $\theta_{12}$ and $\theta_{23}$ as $\theta_{12}$ are found to be around $32.41^{\circ}-33.20^{\circ}$, whereas $\theta_{23}$ are well in higher octant. Here, the values of $\epsilon_{n}$,$\epsilon_{n1}$, $\epsilon_{l}$, and $ \epsilon_{l}^{*}$ are considered very small to have small but necessary deviations from the earlier model while maintaining the correct neutrino mixing texture. These values are also considered to study the required bias needed for lifting the vacua degeneracy.  

\subsection{Gravitational Wave Spectrum from Domain Walls Collision }

To analyse the gravitational wave Spectrum arising from domain wall annihilation, we calculate the $V_{bias}$ for our total potential containing flavon mixing terms and $V_{loop}(\phi,\chi)$. Since the potential has many coupling constants and varying them randomly would give many minima, the values of the coupling constants are fixed in such a way that they can give enough bias to lift the vacua degeneracy and to reproduce $f_{\sigma}$ in the order $\mathcal{O}(1)$. The spectrum of the gravitational waves emitted at time t as a function of frequency $f$ is calculated by solving 
\begin{equation}
\Omega h^2(f,t)=\frac{h^2}{\rho_{c}(t)}\frac{d\rho_{GW}(t)}{d  \ \text{ln}f}
\end{equation} 
To study the red shifting to the present day, one can calculate the peak amplitude and the peak frequency using
\begin{align}
\Omega h^2 |_{peak}& \approx 10^{-67}\left(\frac{f^{4}_{\sigma}}{\epsilon_{b}^2}\right)\left( \frac{v}{\text{TeV}}\right)^4 
\\
f_{peak} &\approx  3 \times 10^3 \ \text{Hz}\left(\frac{\epsilon_{b}v}{f_{\sigma} \text{TeV}}\right)^{\frac{1}{2}}, 
\end{align}
\begin{figure}[h]
\centering
\subfigure[]{
\includegraphics[width=.40\textwidth]{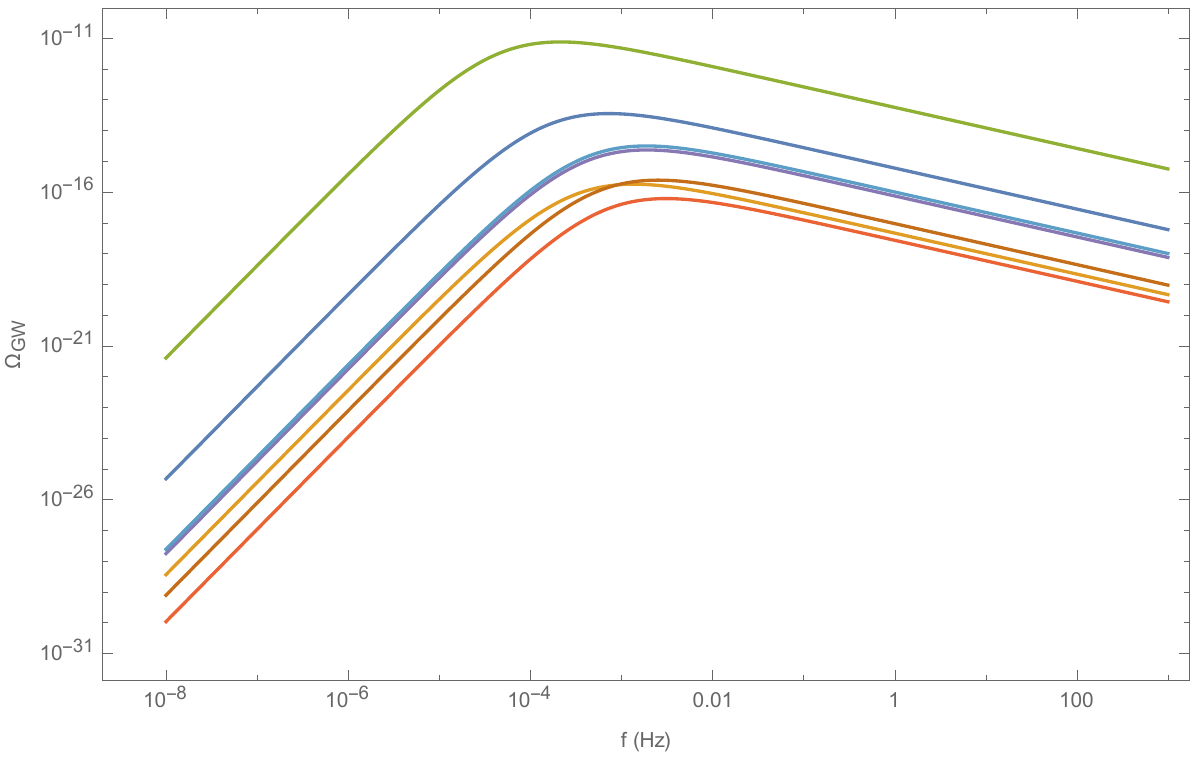}}
\quad
\subfigure[]{
\includegraphics[width=.40\textwidth]{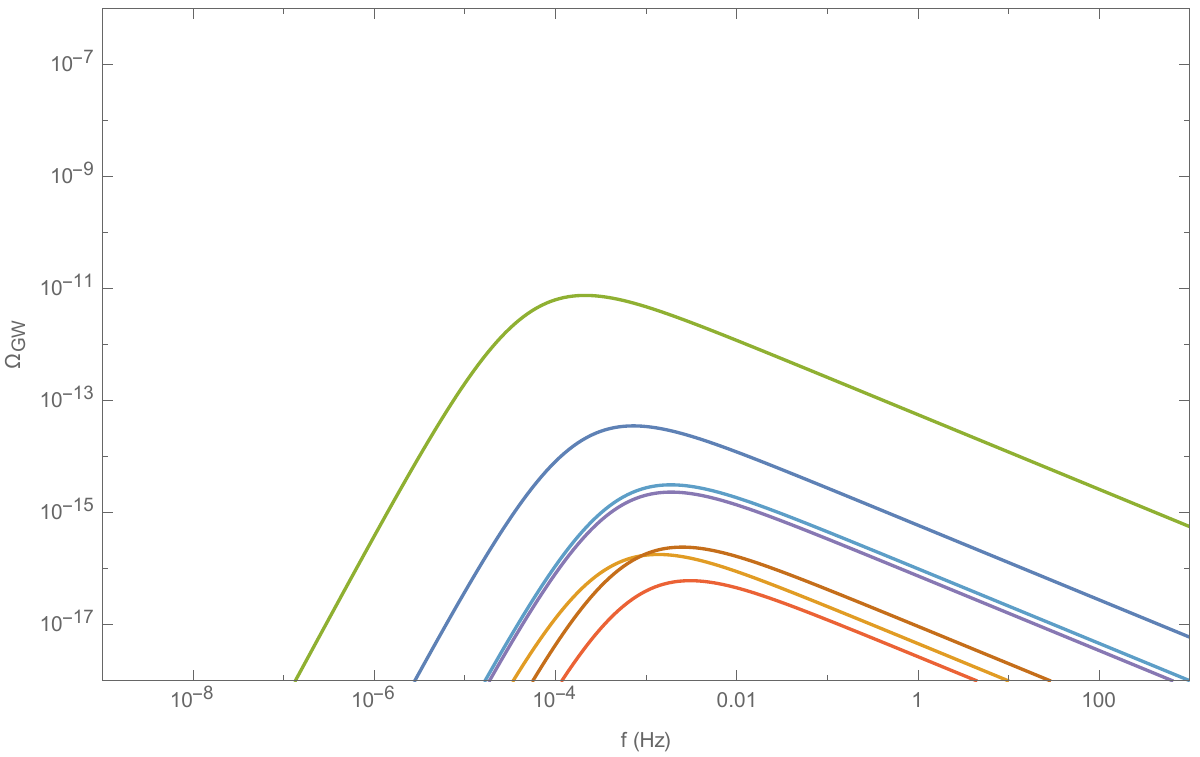}}
\caption{ (a). Gravitational Waves Spectrum for the calculated $f_{sigma}$ and $\epsilon_{b}$ at the vevs of $\phi_{1}$, $\phi_{2}$ and $\phi_{3}$ at $10^4$ TeV. (b). Plots for the part of the spectrum that can be detected by LISA \cite{amaroseoane2017laserinterferometerspaceantenna}, LIGO \cite{Abbott_2019}, PTA \cite{Lentati_2015}, DECIGO/BBO \cite{Seto_2001,Corbin_2006}.}
\label{fg} 
\end{figure}

\begin{table}[ht]
\centering
\begin{tabular}{|c|c|c|c|}
\hline
$\Delta V$ &
$f_{\sigma}$ &
$f_{\rm peak}\,(\mathrm{Hz})$ &
$\Omega_{\rm GW}h^2\big|_{\rm peak}$ \\
\hline
0.00248115 & 1.50882 & $1.21655\times10^{-4}$ & $8.41856\times10^{-14}$ \\
\hline
0.00218520 & 0.377730 & $2.28179\times10^{-4}$ & $4.26329\times10^{-16}$ \\
\hline
0.000265444 & 1.88655 & $3.55856\times10^{-5}$ & $1.79773\times10^{-11}$ \\
\hline
0.0329475 & 1.12019 & $5.14502\times10^{-4}$ & $1.45052\times10^{-16}$ \\
\hline
0.0293781 & 2.62901 & $3.17130\times10^{-4}$ & $5.53504\times10^{-15}$ \\
\hline
0.0295469 & 1.49792 & $4.21340\times10^{-4}$ & $5.76676\times10^{-16}$ \\
\hline
0.0330886 & 3.00674 & $3.14711\times10^{-4}$ & $7.46494\times10^{-15}$ \\
\hline
\end{tabular}
\caption{Calculated domain wall gravitational wave parameters for the spectrum shown in Fig \ref{fg}. Here, the vevs of the flavons are found to be around $\phi_{2}\approx -2\times 10^4$, $\phi_{1}\approx -1.5\times 10^4$, and $\chi\approx 1.4 \times 10^4$. }
\label{gwresults}
\end{table}
The gravitational spectrum is plotted by assuming $\Omega h^2\propto f^3$ for $f<f_{peak}$ and $\Omega h^2 \propto f^{-1}$ for $f>f_{peak}$ motivated by \cite{Hiramatsu_2014}. Fig.\ref{fg}(a) shows the plots for the wave spectrum with different calculated  $f_{\sigma}$. All the values of $f_{\sigma}$ are on the order of $\mathcal{O}(1)$ or lower to have a realistic model, and each spectrum has a different calculated value of $\epsilon_{b}$.  Fig.\ref{fg}(b) shows the part of the GW spectrum that may be detected at present and near-future gravitational wave experiments if $v= 10^4 \text{ TeV}$. The calculated domain wall gravitational wave parameters for the spectrum are summarised in Table \ref{gwresults}.

\section{Summary and Conclusion}
 
 We have presented two simple $A_{4}\times Z_{4}$ models, with and without vev correction, to study neutrino masses and mixing and the GW spectrum due to domain wall annihilation.  The total potential of flavons containing both self-interaction terms and mixed terms is studied to lift vacuum degeneracy of the discrete symmetry group. The mixed terms lift the vacuum degeneracy, correct the vevs, and also provide a slightly different neutrino mixing pattern from the neutrino mixing pattern without them. The neutrino oscillation parameters calculated from both models are in good agreement with current oscillation data. However, the second model shows some interesting findings of $\theta_{12}$ in the range $32.41^{\circ}-33.20^{\circ}$ in opposition to the first model, where they circle around $35.70^{\circ}$. Another interesting feature is that the second model prefers a higher octant in contrast to the first model, where the values of $\theta_{23}$ are spread throughout the current $3\sigma$ range. Further, the GW spectrum that could arise from the domain wall collision is also plotted by calculating the bias and the $f_{sigma}$ from the model that considers the contribution of mixed terms in the potential along with $V_{loop}(\phi_{2},\chi)$. The peak amplitude and the peak frequency for the GW spectrum are also calculated, and they could be detected by present and near-future experiments if the flavon vevs are in $10^4$ TeV.

\section*{Acknowledgements}
One of us (VP)  wishes to thank  Council of Scientific $\&$ Industrial Research (CSIR), Government of India for providing CSIR Research Associate Fellowship. 
\appendix
\section{ $A_{4}$ Group}
$A_{4}$ is the even permutation group of 4 objects with $\frac{4!}{2}$ elements. It has four irreducible representations, namely 1, $1'$, $1''$ and 3. All the elements of the group can be generated by two elements S and T. The generators S and T satisfy the relation,
\begin{equation}
S^{2}=(ST)^3=T^3=1.
\end{equation}
The multiplication rules of any two irreducible representations  under $A_{4}$  are given by
$$ 3\otimes 3=1\oplus 1^{'}\oplus 1^{''}\oplus 3_{S}\oplus  3_{A}$$ 
\begin{equation}
\begin{array}{ccc}
1\otimes1=1 \hspace{2cm}& 1'\otimes1'=1''\\
1"\otimes1''=1' \hspace{2cm} &  1'\otimes1''=1\\
3\otimes 1/1'/1''= 3\hspace{2cm}&1/1'/1''\otimes3=3
\end{array}.
\end{equation}
In the AF basis, 
\begin{align}
1=&a_{1}b_{1}+a_{2}b_{3}+a_{3}b_{2} \nonumber \\
1^{'}=&a_{3}b_{3}+a_{1}b_{2}+a_{2}b_{1} \nonumber \\
1^{''}=&a_{2}b_{2}+a_{1}b_{3}+a_{3}b_{1} \nonumber \\
3_{S}=&\begin{pmatrix}
2a_{1}b_{1}-a_{2}b_{3}-a_{3}b_{2}\\
2a_{3}b_{3}-a_{1}b_{2}-a_{2}b_{1}\\
2a_{2}b_{2}-a_{1}b_{3}-a_{3}b_{1}
\end{pmatrix} \nonumber \\
3_{A}=&\begin{pmatrix}
a_{2}b_{3}-a_{3}b_{2}\\
a_{1}b_{2}-a_{2}b_{1}\\
a_{3}b_{1}-a_{1}b_{3}
\end{pmatrix}
\label{af}
\end{align}

In the MR basis, 
\begin{align}
1=&a_{1}b_{1}+a_{2}b_{2}+a_{3}b_{3} \nonumber \\
1^{'}=&a_{1}b_{1}+\omega a_{1}b_{2}+\omega^{2}a_{3}b_{3} \nonumber \\
1^{''}=&a_{1}b_{1}+\omega^2 a_{1}b_{2}+\omega a_{3}b_{3} \nonumber \\
3_{S}=&\frac{\sqrt{3}}{2}\begin{pmatrix}
a_{2}b_{3}+a_{3}b_{2}\\
a_{3}b_{1}+a_{1}b_{3}\\
a_{1}b_{2}+a_{2}b_{1}
\end{pmatrix}\nonumber \\
3_{S}=&\frac{i}{2}\begin{pmatrix}
a_{2}b_{3}-a_{3}b_{2}\\
a_{3}b_{1}-a_{1}b_{3}\\
a_{1}b_{2}-a_{2}b_{1}
\end{pmatrix}
\end{align}

\section{Solution to Scalar Potential in AF basis \label{b2}}

The scalar potentials given in Eqn (\ref{e1}) and Eqn (\ref{e9}) can be solved in AF basis as given below

\begin{align}
\left\langle\phi_{1}\right\rangle_{-}=\left\lbrace \begin{pmatrix}
\sqrt{3}\\0\\0
\end{pmatrix}, \ \begin{pmatrix}
-\frac{1}{\sqrt{3}}\\ \frac{2}{\sqrt{3}}\omega^2 \\ \frac{2}{\sqrt{3}}\omega
\end{pmatrix},\ 
\begin{pmatrix}
-\frac{1}{\sqrt{3}}\\ \frac{2}{\sqrt{3}}\omega \\ \frac{2}{\sqrt{3}}\omega^2
\end{pmatrix},\ 
\begin{pmatrix}
-\frac{1}{\sqrt{3}}\\ \frac{2}{\sqrt{3}} \\ \frac{2}{\sqrt{3}}
\end{pmatrix} \right\rbrace v_{\phi_{1}^{-}}
\label{p1}
\end{align}
and 
\begin{align}
\left\langle\phi_{1}\right\rangle_{+}=\left\lbrace \begin{pmatrix}
-\sqrt{3}\\0\\0
\end{pmatrix}, \ \begin{pmatrix}
\frac{1}{\sqrt{3}}\\ -\frac{2}{\sqrt{3}}\omega^2 \\ -\frac{2}{\sqrt{3}}\omega
\end{pmatrix},\ 
\begin{pmatrix}
\frac{1}{\sqrt{3}}\\ -\frac{2}{\sqrt{3}}\omega \\ -\frac{2}{\sqrt{3}}\omega^2
\end{pmatrix},\ 
\begin{pmatrix}
\frac{1}{\sqrt{3}}\\ -\frac{2}{\sqrt{3}} \\ -\frac{2}{\sqrt{3}}
\end{pmatrix} \right\rbrace v_{\phi_{1}^{+}}
\label{p2}
\end{align}
and
\begin{align}
\left\langle\phi_{2}\right\rangle=\left\lbrace \begin{pmatrix}
1\\1\\1
\end{pmatrix}, \ \begin{pmatrix}
1\\ \omega \\ \omega^2
\end{pmatrix},\ \begin{pmatrix}
1\\ \omega^2 \\ \omega
\end{pmatrix}\right\rbrace
\label{p3}
\end{align}
\bibliographystyle{unsrtnat}
\bibliography{refnew}

\begin{thebibliography}{36}
\providecommand{\natexlab}[1]{#1}
\providecommand{\url}[1]{\texttt{#1}}
\expandafter\ifx\csname urlstyle\endcsname\relax
  \providecommand{\doi}[1]{doi: #1}\else
  \providecommand{\doi}{doi: \begingroup \urlstyle{rm}\Url}\fi

\bibitem[Eguchi et~al.(2003)]{KamLAND:2002uet}
K.~Eguchi et~al.
\newblock {First results from KamLAND: Evidence for reactor anti-neutrino
  disappearance}.
\newblock \emph{Phys. Rev. Lett.}, 90:\penalty0 021802, 2003.
\newblock \doi{10.1103/PhysRevLett.90.021802}.

\bibitem[Ahmad et~al.(2002)]{SNO:2002tuh}
Q.~R. Ahmad et~al.
\newblock {Direct evidence for neutrino flavor transformation from neutral
  current interactions in the Sudbury Neutrino Observatory}.
\newblock \emph{Phys. Rev. Lett.}, 89:\penalty0 011301, 2002.
\newblock \doi{10.1103/PhysRevLett.89.011301}.

\bibitem[Fukuda et~al.(1998)]{Super-Kamiokande:1998kpq}
Y.~Fukuda et~al.
\newblock {Evidence for oscillation of atmospheric neutrinos}.
\newblock \emph{Phys. Rev. Lett.}, 81:\penalty0 1562--1567, 1998.
\newblock \doi{10.1103/PhysRevLett.81.1562}.

\bibitem[Abe et~al.(2012)]{DoubleChooz:2011ymz}
Y.~Abe et~al.
\newblock {Indication of Reactor $\bar{\nu}_e$ Disappearance in the Double
  Chooz Experiment}.
\newblock \emph{Phys. Rev. Lett.}, 108:\penalty0 131801, 2012.
\newblock \doi{10.1103/PhysRevLett.108.131801}.

\bibitem[Lasserre et~al.(2012)Lasserre, Mention, Cribier, Collin, Durand,
  Fischer, Gaffiot, Lhuillier, Letourneau, and Vivier]{Lasserre:2012ax}
Thierry Lasserre, Guillaume Mention, Michel Cribier, Antoine Collin, Vincent
  Durand, Vincent Fischer, Jonathan Gaffiot, David Lhuillier, Alain Letourneau,
  and Matthieu Vivier.
\newblock {Comment on Phys. Rev. Lett. 108, 191802 (2012): 'Observation of
  Reactor Electron Antineutrino Disappearance in the RENO Experiment'}.
\newblock 5 2012.

\bibitem[Ling(2013)]{Ling:2013fta}
Jiajie Ling.
\newblock {Observation of electron-antineutrino disappearance at Daya Bay}.
\newblock \emph{AIP Conf. Proc.}, 1560\penalty0 (1):\penalty0 206--210, 2013.
\newblock \doi{10.1063/1.4826754}.

\bibitem[McDonald(2016)]{McDonald:2016ixn}
Arthur~B. McDonald.
\newblock {Nobel Lecture: The Sudbury Neutrino Observatory: Observation of
  flavor change for solar neutrinos}.
\newblock \emph{Rev. Mod. Phys.}, 88\penalty0 (3):\penalty0 030502, 2016.
\newblock \doi{10.1103/RevModPhys.88.030502}.

\bibitem[Altmannshofer and Greljo(2024)]{Altmannshofer:2024hmr}
Wolfgang Altmannshofer and Admir Greljo.
\newblock {Recent Progress in Flavor Model Building}.
\newblock 12 2024.
\newblock \doi{10.1146/annurev-nucl-121423-100950}.

\bibitem[King and Luhn(2013)]{King:2013eh}
Stephen~F. King and Christoph Luhn.
\newblock {Neutrino Mass and Mixing with Discrete Symmetry}.
\newblock \emph{Rept. Prog. Phys.}, 76:\penalty0 056201, 2013.
\newblock \doi{10.1088/0034-4885/76/5/056201}.

\bibitem[Aslam et~al.(2026)Aslam, Zafar, Aslam, Mirza, Saleem, and
  Hasnaoui]{Aslam:2026zrw}
Muhammad~Waheed Aslam, Abrar~Ahmad Zafar, Muhammad~Naeem Aslam, Arifa Mirza,
  Salman Saleem, and Abdelhalim Hasnaoui.
\newblock {Revisiting a novel A4 scenario for optimizing neutrino masses with
  AI-Based algorithm}.
\newblock \emph{Phys. Lett. B}, 874:\penalty0 140247, 2026.
\newblock \doi{10.1016/j.physletb.2026.140247}.

\bibitem[Chuli{\'a} and Kumar(2026)]{Chulia:2025nsa}
Salvador~Centelles Chuli{\'a} and Ranjeet Kumar.
\newblock {Minimal \(A_{4}\) type-II seesaw realization of testable neutrino
  mass sum rules}.
\newblock \emph{Phys. Rev. D}, 113\penalty0 (5):\penalty0 055023, 2026.
\newblock \doi{10.1103/ndvc-1vpl}.

\bibitem[C\'arcamo~Hern\'andez and
  de~Medeiros~Varzielas(2022)]{CarcamoHernandez:2022bka}
A.~E. C\'arcamo~Hern\'andez and Ivo de~Medeiros~Varzielas.
\newblock {An A5 inverse seesaw model with perturbed golden ratio mixing}.
\newblock \emph{Nucl. Phys. B}, 985:\penalty0 116031, 2022.
\newblock \doi{10.1016/j.nuclphysb.2022.116031}.

\bibitem[Thapa et~al.(2023)Thapa, Barman, Bora, and Francis]{Thapa_2023}
Bikash Thapa, Sunita Barman, Sompriti Bora, and N.~K. Francis.
\newblock A minimal inverse seesaw model with s4 flavour symmetry.
\newblock \emph{Journal of High Energy Physics}, 2023\penalty0 (11), November
  2023.
\newblock ISSN 1029-8479.
\newblock \doi{10.1007/jhep11(2023)154}.
\newblock URL \url{http://dx.doi.org/10.1007/JHEP11(2023)154}.

\bibitem[Pathak and Das(2026)]{Pathak:2026ezl}
Gourab Pathak and Mrinal~Kumar Das.
\newblock {Froggatt-Nielsen like mechanism in inverse seesaw using modular
  symmetry}.
\newblock \emph{J. Subatomic Part. Cosmol.}, 6:\penalty0 100432, 2026.
\newblock \doi{10.1016/j.jspc.2026.100432}.

\bibitem[Nomura and Okada(2025)]{Nomura:2023kwz}
Takaaki Nomura and Hiroshi Okada.
\newblock {Quark and lepton model with flavor specific dark matter and muon \(g
  - 2\) in modular \(A_{4}\) and hidden \(U(1)\) symmetries}.
\newblock \emph{Phys. Dark Univ.}, 49:\penalty0 101986, 2025.
\newblock \doi{10.1016/j.dark.2025.101986}.

\bibitem[de~Medeiros~Varzielas and
  Louren\c{c}o(2022)]{deMedeirosVarzielas:2022ihu}
Ivo de~Medeiros~Varzielas and Jo\~ao Louren\c{c}o.
\newblock {Two A5 modular symmetries for Golden Ratio 2 mixing}.
\newblock \emph{Nucl. Phys. B}, 984:\penalty0 115974, 2022.
\newblock \doi{10.1016/j.nuclphysb.2022.115974}.

\bibitem[Kalita and Patgiri(2026)]{Kalita:2026zhm}
Raktima Kalita and Mahadev Patgiri.
\newblock {Modular S4 invariant left-right symmetric linear seesaw neutrino
  models}.
\newblock \emph{Phys. Dark Univ.}, 52:\penalty0 102330, 2026.
\newblock \doi{10.1016/j.dark.2026.102330}.

\bibitem[Heinrich et~al.(2019)Heinrich, Schulz, Turner, and
  Zhou]{Heinrich_2019}
Lukas Heinrich, Holger Schulz, Jessica Turner, and Ye-Ling Zhou.
\newblock Constraining a4 leptonic flavour model parameters at colliders and
  beyond.
\newblock \emph{Journal of High Energy Physics}, 2019\penalty0 (4), April 2019.
\newblock ISSN 1029-8479.
\newblock \doi{10.1007/jhep04(2019)144}.
\newblock URL \url{http://dx.doi.org/10.1007/JHEP04(2019)144}.

\bibitem[Gouttenoire et~al.(2025)Gouttenoire, King, Roshan, Wang, White, and
  Yamazaki]{Gouttenoire_2025}
Yann Gouttenoire, Stephen~F. King, Rishav Roshan, Xin Wang, Graham White, and
  Masahito Yamazaki.
\newblock Cosmological consequences of domain walls biased by quantum gravity.
\newblock \emph{Physical Review D}, 112\penalty0 (7), October 2025.
\newblock ISSN 2470-0029.
\newblock \doi{10.1103/7zmx-v16z}.
\newblock URL \url{http://dx.doi.org/10.1103/7zmx-v16z}.

\bibitem[King et~al.(2024)King, Leontaris, and Zhou]{King_2024}
Stephen~F. King, George~K. Leontaris, and Ye-Ling Zhou.
\newblock Flipped su(5): unification, proton decay, fermion masses and
  gravitational waves.
\newblock \emph{Journal of High Energy Physics}, 2024\penalty0 (3), March 2024.
\newblock ISSN 1029-8479.
\newblock \doi{10.1007/jhep03(2024)006}.
\newblock URL \url{http://dx.doi.org/10.1007/JHEP03(2024)006}.

\bibitem[Jueid et~al.(2023)Jueid, Loualidi, Nasri, and
  Ouahid]{jueid2023cosmologicaldomainwallsbreaking}
Adil Jueid, Mohamed~Amin Loualidi, Salah Nasri, and Mohamed~Amine Ouahid.
\newblock Cosmological domain walls from the breaking of $\mathbf{S_4}$ flavor
  symmetry, 2023.
\newblock URL \url{https://arxiv.org/abs/2312.04388}.

\bibitem[Zeldovich et~al.(1974)Zeldovich, Kobzarev, and Okun]{Zeldovich:1974uw}
Ya.~B. Zeldovich, I.~Yu. Kobzarev, and L.~B. Okun.
\newblock {Cosmological Consequences of the Spontaneous Breakdown of Discrete
  Symmetry}.
\newblock \emph{Zh. Eksp. Teor. Fiz.}, 67:\penalty0 3--11, 1974.

\bibitem[Gelmini et~al.(2021)Gelmini, Pascoli, Vitagliano, and
  Zhou]{Gelmini:2020bqg}
Graciela~B. Gelmini, Silvia Pascoli, Edoardo Vitagliano, and Ye-Ling Zhou.
\newblock {Gravitational wave signatures from discrete flavor symmetries}.
\newblock \emph{JCAP}, 02:\penalty0 032, 2021.
\newblock \doi{10.1088/1475-7516/2021/02/032}.

\bibitem[Chigusa and Nakayama(2019)]{CHIGUSA2019249}
So~Chigusa and Kazunori Nakayama.
\newblock Anomalous discrete flavor symmetry and domain wall problem.
\newblock \emph{Physics Letters B}, 788:\penalty0 249--255, 2019.
\newblock ISSN 0370-2693.
\newblock \doi{https://doi.org/10.1016/j.physletb.2018.11.027}.
\newblock URL
  \url{https://www.sciencedirect.com/science/article/pii/S0370269318308670}.

\bibitem[Chen et~al.(2026)Chen, Matias, and
  Moffett-Smith]{chen2026gravitationalwavesa4neutrino}
Mu-Chun Chen, Harold~J. Matias, and Cameron Moffett-Smith.
\newblock Gravitational waves in an a4 neutrino mass model, 2026.
\newblock URL \url{https://arxiv.org/abs/2601.14394}.

\bibitem[Preskill et~al.(1991)Preskill, Trivedi, Wilczek, and
  Wise]{PRESKILL1991207}
John Preskill, Sandip~P. Trivedi, Frank Wilczek, and Mark~B. Wise.
\newblock Cosmology and broken discrete symmetry.
\newblock \emph{Nuclear Physics B}, 363\penalty0 (1):\penalty0 207--220, 1991.
\newblock ISSN 0550-3213.
\newblock \doi{https://doi.org/10.1016/0550-3213(91)90241-O}.
\newblock URL
  \url{https://www.sciencedirect.com/science/article/pii/055032139190241O}.

\bibitem[Gelmini et~al.(1989)Gelmini, Gleiser, and Kolb]{PhysRevD.39.1558}
Graciela~B. Gelmini, Marcelo Gleiser, and Edward~W. Kolb.
\newblock Cosmology of biased discrete symmetry breaking.
\newblock \emph{Phys. Rev. D}, 39:\penalty0 1558--1566, Mar 1989.
\newblock \doi{10.1103/PhysRevD.39.1558}.
\newblock URL \url{https://link.aps.org/doi/10.1103/PhysRevD.39.1558}.

\bibitem[Hiramatsu et~al.(2014)Hiramatsu, Kawasaki, and
  Saikawa]{Hiramatsu_2014}
Takashi Hiramatsu, Masahiro Kawasaki, and Ken’ichi Saikawa.
\newblock On the estimation of gravitational wave spectrum from cosmic domain
  walls.
\newblock \emph{Journal of Cosmology and Astroparticle Physics}, 2014\penalty0
  (02):\penalty0 031–031, February 2014.
\newblock ISSN 1475-7516.
\newblock \doi{10.1088/1475-7516/2014/02/031}.
\newblock URL \url{http://dx.doi.org/10.1088/1475-7516/2014/02/031}.

\bibitem[Pascoli and Zhou(2016{\natexlab{a}})]{Pascoli_2016}
Silvia Pascoli and Ye-Ling Zhou.
\newblock The role of flavon cross couplings in leptonic flavour mixing.
\newblock \emph{Journal of High Energy Physics}, 2016\penalty0 (6),
  2016{\natexlab{a}}.
\newblock ISSN 1029-8479.
\newblock \doi{10.1007/jhep06(2016)073}.
\newblock URL \url{http://dx.doi.org/10.1007/JHEP06(2016)073}.

\bibitem[Pascoli and Zhou(2016{\natexlab{b}})]{Pascoli_2016a}
Silvia Pascoli and Ye-Ling Zhou.
\newblock Flavon-induced connections between lepton flavour mixing and charged
  lepton flavour violation processes.
\newblock \emph{Journal of High Energy Physics}, 2016\penalty0 (10), October
  2016{\natexlab{b}}.
\newblock ISSN 1029-8479.
\newblock \doi{10.1007/jhep10(2016)145}.
\newblock URL \url{http://dx.doi.org/10.1007/JHEP10(2016)145}.

\bibitem[Esteban et~al.(2024)Esteban, Gonzalez-Garcia, Maltoni, Martinez-Soler,
  Pinheiro, and Schwetz]{Esteban_2024}
Ivan Esteban, M.~C. Gonzalez-Garcia, Michele Maltoni, Ivan Martinez-Soler,
  João~Paulo Pinheiro, and Thomas Schwetz.
\newblock Nufit-6.0: updated global analysis of three-flavor neutrino
  oscillations.
\newblock \emph{Journal of High Energy Physics}, 2024\penalty0 (12), December
  2024.
\newblock ISSN 1029-8479.
\newblock \doi{10.1007/jhep12(2024)216}.
\newblock URL \url{http://dx.doi.org/10.1007/JHEP12(2024)216}.

\bibitem[et~al.(2017)]{amaroseoane2017laserinterferometerspaceantenna}
Pau Amaro-Seoane et~al.
\newblock Laser interferometer space antenna, 2017.
\newblock URL \url{https://arxiv.org/abs/1702.00786}.

\bibitem[et~al.(2019)]{Abbott_2019}
B.~Abbott et~al.
\newblock Search for the isotropic stochastic background using data from
  advanced ligo’s second observing run.
\newblock \emph{Physical Review D}, 100\penalty0 (6), 2019.
\newblock ISSN 2470-0029.
\newblock \doi{10.1103/physrevd.100.061101}.
\newblock URL \url{http://dx.doi.org/10.1103/PhysRevD.100.061101}.

\bibitem[et~al.(2015)]{Lentati_2015}
L.~Lentati et~al.
\newblock European pulsar timing array limits on an isotropic stochastic
  gravitational-wave background.
\newblock \emph{Monthly Notices of the Royal Astronomical Society},
  453\penalty0 (3):\penalty0 2577–2599, August 2015.
\newblock ISSN 1365-2966.
\newblock \doi{10.1093/mnras/stv1538}.
\newblock URL \url{http://dx.doi.org/10.1093/mnras/stv1538}.

\bibitem[Seto et~al.(2001)Seto, Kawamura, and Nakamura]{Seto_2001}
Naoki Seto, Seiji Kawamura, and Takashi Nakamura.
\newblock Possibility of direct measurement of the acceleration of the universe
  using 0.1 hz band laser interferometer gravitational wave antenna in space.
\newblock \emph{Physical Review Letters}, 87\penalty0 (22), November 2001.
\newblock ISSN 1079-7114.
\newblock \doi{10.1103/physrevlett.87.221103}.
\newblock URL \url{http://dx.doi.org/10.1103/PhysRevLett.87.221103}.

\bibitem[Corbin and Cornish(2006)]{Corbin_2006}
Vincent Corbin and Neil~J Cornish.
\newblock Detecting the cosmic gravitational wave background with the big bang
  observer.
\newblock \emph{Classical and Quantum Gravity}, 23\penalty0 (7):\penalty0
  2435–2446, March 2006.
\newblock ISSN 1361-6382.
\newblock \doi{10.1088/0264-9381/23/7/014}.
\newblock URL \url{http://dx.doi.org/10.1088/0264-9381/23/7/014}.

\end{thebibliography}
\end{document}